
\def\unlockat{\catcode`\@=11} 
\def\lockat{\catcode`\@=12}   
\unlockat
\def\m@ssage{\immediate\write16}  \m@ssage{}
\m@ssage{if two columns per page, type any character now}
\m@ssage{otherwise, or if confused, hit return.}
\def\d@f@ult{}
\newif\ifdoubl@col
\endlinechar=-1  
\read-1 to\@nswer
\endlinechar=13
\ifx\@nswer\d@f@ult\doubl@colfalse\m@ssage{Single column.}
 \else\doubl@coltrue\m@ssage{Two columns per page.  Use landscape mode
when printing.}\fi
\newdimen\fullhsize \newbox\leftpage
\newif\ifl@stp@geempty \l@stp@geemptyfalse
\ifdoubl@col

 \fullhsize=25cm \hsize=12cm \vsize=18cm
 \let\l@r=L
 \output={\almostshipout{\leftline{\vbox{
           \makeheadline\pagebody\makefootline}}}
         \advancepageno}
 \def\almostshipout#1{%
     \if L\l@r \count1=1
       \message{[\the\count0.\the\count1]}
       \global\setbox\leftpage=#1 \global\let\l@r=R
     \else \count1=2
       \shipout\vbox{\hbox to \fullhsize{\box\leftpage\hfil #1}}
       \global\let\l@r=L\fi}
\hoffset=-1.5mm \voffset=-15.5mm
\baselineskip=12pt
\else
\magnification=\magstep1  
\vsize=8.4truein
\voffset=0.5truein
\hoffset=0.35truein   
\hsize=6.3truein     
\baselineskip=16pt
\fi
\outer\def\bye{\par\vfill\supereject
    \ifdoubl@col\if R\l@r \l@stp@geemptytrue\null\vfill\eject\fi\fi
    \end}
\lockat
\nonfrenchspacing    
\interlinepenalty=10  
\overfullrule=0pt

\font\grosss=cmr7 scaled \magstep4
\font\gross=cmr10 scaled \magstep2

\outer\def\unter#1 #2\par{\vskip 0pt plus .1\vsize \penalty-200
     \vskip 0pt plus -.1\vsize \bigskip \vskip \parskip
     \message{#1 #2
       } \vfill\leftline{\gross #1 \bf#2}\nobreak\medskip\noindent}

\def \ccc{{\rm C \kern-5.5pt I \ }}
\def \zzz{{\rm \, Z \!\! Z}}
\def \nnn{{\rm I \! N}}

\def \ccp{{\rm C \kern-5.5pt I\, P}}

\def \mod{{\rm\ mod\ }}

\def \eins{{\rm 1 \kern-2.8pt I }}

\def \lll{{\rm L}}

\def \>>{\rangle}      
\def \tchi{\chi \kern-5.6pt\chi }
\def \tzeta{\zeta \kern-5.6pt\zeta }
\def \tD{\Delta \kern-7.6pt\Delta}
\def \tmu{\mu \kern-5.6pt\mu }
\def \teta{\eta \kern-5.6pt\eta }
\def \txi{\xi \kern-5.0pt\xi }
\def \LL{{\rm Li_2}}

\def \res{{\rm Res}}
\def \emu{{\cal E}_{\tmu,l}}
\def \eml{{\cal E}_{l}}

\def\frac#1/#2{\leavevmode\kern.1em
              \raise.5ex\hbox{$\the\scriptfont0 #1$}\kern-.1em
              /\kern-.05em\lower.4ex\hbox{$\the\scriptfont0 #2$}}

\def \sqr#1#2{{\vcenter{\vbox{\hrule height.#2pt
            \hbox{\vrule width.#2pt height#1pt \kern#1pt
                   \vrule width.#2pt}
            \hrule height.#2pt}}}}

\def \ie{\frenchspacing i.e.\ }
\def \cf{\frenchspacing cf.\ }

\def \bb{{\bf b}}
\def \cc{{\bf c}}
\def \ww{{\bf w}}
\def \vv{{\bf v}}

\def \cef{c_{\rm eff}}
\def \hm{h_{\rm min}}

\def \eg{\frenchspacing e.g.\ }


{\nopagenumbers

\null
\line{\hfil preprint BONN-HE-93-24}
\line{\hfil hep-th/9307056}
\line{\hfil July 1993}
\vfill
\centerline {\grosss Dilogarithm identities, fusion rules}
\medskip
\centerline {\grosss and structure constants of CFTs}
\bigskip
\centerline{\phantom{mmmmmmmmmmmmm}Michael Terhoeven}
\bigskip
\bigskip
\centerline{Physikalisches Institut}
\centerline{der}
\centerline{Rheinischen Friedrich-Wilhelms Universit\"at Bonn}
\centerline{Nussallee 12}
\centerline{D-53115 Bonn}
\smallskip
\centerline{unp044 at ibm.rhrz.uni-bonn.de}
\vfill
\centerline{\bf Abstract\phantom{mmmmmmmmmmmmmmmm}}
\medskip
Recently dilogarithm identities have made their appearance in the
physics literature. These identities seem to allow to calculate
structure constants like, in particular, the effective central
charge of certain conformal field theories from their fusion rules.
In Nahm, Recknagel, Terhoeven (1992) a proof of identities of
this type was given by considering the asymptotics of character
functions in the so-called Rogers-Ramanujan sum form and
comparing with the asymptotics predicted by modular covariance.
Refining the argument, we obtain {\it the general connection
of quantum dimensions of certain conformal field theories
to the arguments of the dilogarithm function} in the identities
in question and {\it an infinite set of consistency conditions
on the parameters of Rogers-Ramanujan type partitions for them
to be modular covariant}.
\bigskip
\vfill
\eject }

\pageno=1

\unter 1 Introduction and Overview

Recently, the dilogarithm function has reappeared (Babujan (1983))
in the physics of two-dimensional quantum field theories and
lattice models (Bazhanov, Reshetikhin (1989), Klassen, Melzer
(1990), Kl\"umper, Pearce (1991), Zamolodchikov (1990)).
In Nahm, Recknagel, Terhoeven (1992) a connection of fusion
rules to thermodynamic Bethe Ansatz (TBA) type equations was
conjectured. In this paper, we make this connection precise
in the framework of CFT.

Crucial in the latter approach of the subject is to (re)write
the characters of the conformal field theory (CFT) in the
Rogers-Ramanujan sum form (\cf (1.1) below). In Terhoeven (1992),
Kuniba, Nakanishi, Suzuki (1992) this form was conjectured to
exist for general parafermionic theories. Allowing a slight
generalization, one obtains formulas of similar type for
characters of all unitary Virasoro minimal models
(\cf Dashmahapatra, Kedem, Klassen, McCoy, Melzer (1993) and
references therein, also for the so-called quasi-particle
picture). Thus it seems possible that the class
of theories allowing their characters to be written in a
suitable form is not too far away from the set of all CFTs.
In this paper, we find an infinite set of consistency
conditions to be satisfied by all partitions of the form (1.1),
which is a first step in the classification of this type of
partitions, possibly relevant for a classification of
all rational CFTs.

In the following, we consider partition functions of the
Rogers-Ramanujan (or Euler) sum form
$$
\eqalign{
  Z(q) &=
 \sum_{\matrix{&\scriptstyle m_1,\ldots ,m_n\in\zzz_{\geq 0}\cr
               &\scriptstyle c\cdot {\bf m} = \gamma\mod 1\cr}}
           { q^{{\bf m}B{\bf m}^t + {\bf b\cdot m} + \beta }
                 \over (q)_{\bf m}}       \cr
      &= \sum_{l\in\zzz_K} {1\over K}
 \sum_{ m_1,\ldots ,m_n\in\zzz_{\geq 0}}
           { q^{{\bf m}B{\bf m}^t + {\bf b\cdot m} + \beta }
                 \over (q)_{\bf m}}\ e^{-2\pi il({\bf c\cdot m}-\gamma)},
 \cr }
\eqno(1.1)
$$
where $(q)_{\bf m}=(q)_{m_1}\ldots (q)_{m_n}$,
$(q)_m = (1-q)(1-q^2)\ldots(1-q^m)$, $B_{ij}$, $b_i$, $c_i$,
$\beta$ are rational numbers,
$\gamma \in {\zzz_K}/K$ ($K\in\nnn$) and
$B$ is symmetric and positive definite.
Let $q = e^{2\pi i\tau}=e^{-2\pi t}$ and
$\tilde q = e^{-2\pi i/\tau}=e^{-2\pi /t}$,
thus $1/\log(q) = \log(\tilde q)/4\pi^2$ and in the limit
$t=-i\tau \searrow 0$ we have $q \nearrow 1$.

More accurately, we consider the asymptotic of $(1.1)$ in the above
limit. Refining the argument of Nahm, Recknagel, Terhoeven (1992)
and requiring (1.1) to be modular covariant,
we obtain not only a dilogarithm identity for the lowest
energy state in the spectrum. More than that: our main results
are {\it the connection of fusion rules to TBA-type equations}
(meaning a formula relating quantum dimensions and a certain
solution of the TBA-type equations, the arguments of the
dilogarithm function in the formula for the effective central
charge)
and {\it an infinite set of consistency equations restricting
the possible choices of $B$, $\bb$ and $\beta$} (\cf the discussion
in section four).

\medskip
\noindent
The outline of the paper is as follows:

In the rather technical second section we study the asymptotics of
(1.1) for $q \nearrow 1$ using an integral representation and
integral transform of the partition function following Meinardus
(1954) and a higher order saddle point approximation, preferable
diagrammatically (\`a la Feynman).

In the third section we compare the partition asymptotics calculated
explicitly with the one predicted by modular covariance, and
thus find the results mentioned above.

Finally, we outline future work and as a short application
present a classification of modular covariant partitions of the
Rogers-Ramanujan form in the case rank $B =n=1$, which is found
from the first few consistency equations in this particular case.

\medskip 
\unter 2 Partition asymptotics

In this section, we will extract the first term of the modular
transform of $Z(q)$ -- a power series in $\tilde q$ -- from the
asymptotic behavior of (1.1) when $q\nearrow 1$.

\medskip
Following Meinardus (1954), we write (1.1) in an integral
representation using {\it Cauchy's theorem}
$$
Z(q) =\oint\ \prod_j{dw_j\over 2\pi i\ w_j}\ {q^{\beta}\over K}\
     \big( \sum_{\matrix{\scriptstyle \tmu\in\zzz^n\cr
                         \scriptstyle l\in\zzz_K  \cr}}
     q^{\tmu B \tmu^t }\ e^{-2\pi i l(\tmu\cdot\cc -\gamma)} \
     \prod_j  w_j^{-\mu_j}  \big)\
     \prod_j \Big( \sum_{m_j\geq 0} {q^{ b_j m_j } \over (q)_{m_j}}
                   w_j^{m_j} \Big) .
\eqno(2.1)
$$
On the first bracket inside the multiple integral in (2.1)
we apply the well-known {\it Jacobi inversion formula} (Gunning (1962))
($|q|<1$)
$$
\eqalign{
    &\qquad \sum_{\tmu\in\zzz^n,\ l\in\zzz_K}
  q^{ \tmu B \tmu}\ e^{-2\pi i l(\tmu\cdot\cc -\gamma)} \ \ww^{\tmu}\cr
       &= {1\over \sqrt{\det(2B)\ t^n }}\
          \sum_{l\in\zzz_K}\ e^{2\pi i l\gamma}\
           \sum_{\tmu\in\zzz^n}
          {\tilde q}^{(\tmu+\log(\ww)/2\pi i+l\cc)
              B^{-1}(\tmu+\log(\ww)/2\pi i+l\cc)/4}, \cr}
\eqno(2.2)
$$
where $\ww^{\tmu}=\prod_j w_j^{\mu_j}$.
The terms in the second bracket of (2.1) require some more
explicit calculations. First, we use a relation going back
to {\it Euler} (Andrews (1976)) ($|q|<1, |w|<|q|^{-b}$)
$$
 \sum_{m\geq 0} {q^{ b m } \over (q)_{m}}\ w^m =
  \prod_{n\geq 0} (1-w\ q^{n+b})^{-1} .
\eqno(2.3)
$$
Second, we apply ${\rm id}=\exp\circ\log$ and the Taylor series
expansion of $\log(1-x)$ to obtain
$$
  \exp\big\{ \sum_{r=1}^{\infty} {w^r\over r}
             \sum_{n=0}^{\infty} e^{-2\pi t(n+b)r}
        \big\} .
\eqno(2.4)
$$
Using {\it inverse Mellin transformation} (\cf \eg Davies (1978)),
the exponent in (2.4) can be written as
$$
  {1\over 2\pi i} \int_{3/2-i\infty}^{3/2+i\infty}
       ds\ \Gamma(s)\ D(s+1,w)\ \zeta(s,b)\ (2\pi t)^{-s},
\eqno(2.5)
$$
where $\Gamma(s)$ is the Gamma function with $\res_{s=-n}\Gamma(s)
= {(-)^n\over n!}$ $(n>0)$, $\zeta(s,b)$ is the generalized Riemann
$\zeta$-function $\zeta(s,b)=\sum_{n=0}^{\infty} (n+b)^{-s}$ with
$\res_{s=-n}\zeta(s,b) = - {\varphi_{n+1}(b)\over (n+1)}$ $(n>0)$,
where $\varphi_n(x)$ are the Bernoulli polynomials ($\varphi_0(x)=1$,
$\varphi_1(x)=x-1/2$, $\varphi_2(x)=x^2-x+1/6$, $\ldots$) generated
by ${te^{xt}\over (e^t-1)}=\sum_{n=0}^{\infty} {t^n\over n!}\
\varphi_n(x)$ and $D(s,w)=\sum_{r=1}^{\infty} {w^r\over r^s}$
($|w|<1$).

\noindent
Third, shifting the contour of integration in (2.5) along the
real axis, we pick up one by one the residues of the integrand
at $s=1,0,-1,\ldots\ $, the first coming from the pole of $\zeta$,
the others from poles of $\Gamma$. This gives the following expression
for the exponent of (2.4)
$$
  -{1\over\log(q)} \ \sum_{p=0}^{\infty}\ \log^p(q)\
       {\varphi_p(b) \over p!}\
        (w\partial_w)^p\
  \sum_{r=1}^{\infty} {w^r\over r^2},
\eqno(2.6)
$$
which, using the generating function of the Bernoulli polynomials
and the series definition of the dilogarithm $\LL(w) =
\sum_{r=1}^{\infty} {w^r\over r^2}$, can be written more
compactly as
$$
  - w\partial_w \
  {q^{bw\partial_w}\over q^{w\partial_w}-1}\quad
    \LL (w).
\eqno(2.7)
$$

\noindent
Taking (2.6) and (2.2), we can write (2.1) in the form
$$
 Z(q) =  \oint\
  \prod_j{dw_j\over 2\pi i\ w_j}\
  {1\over K\sqrt{\det(2B)\ t^n}}\
   \sum_{\tmu\in\zzz^n,\ l\in\zzz_K}\  \exp(\emu(q,\ww)),
\eqno(2.8a)
$$
where
$$
\eqalign{
\emu(q,\ww)= &\log(q)\ \beta\ + \ 2\pi il\gamma\ \cr
& + {4\pi^2\over \log(q)}\ {1\over 4} (\tmu+{\log(\ww)\over 2\pi i}+l\cc)
       B^{-1}(\tmu+{\log(\ww)\over 2\pi i}+l\cc)^t \cr
& -{1\over\log(q)} \ \sum_{p=0}^{\infty}\ \log^p(q)\
       \sum_{a=1}^n\ {\varphi_p(b_a) \over p!}\
       (w_a\partial_{w_a})^p\
       \sum_{r_a=1}^{\infty} {w_a^{r_a}\over r_a^2}, \cr }
\eqno(2.8b)
$$
which in lowest order in $\log(q)$ reduces to
$$
 {1\over \log(q)}\Big( 4\pi^2\  { 1\over 4}
        (\tmu+{\log(\ww)\over 2\pi i}+l\cc)B^{-1}
               (\tmu+{\log(\ww)\over 2\pi i}+l\cc)^t
        - \sum_a\ \LL(w_a)   \Big).
\eqno(2.9)
$$
The sum over $\tmu$ together with the contour integral can be
understood as an infinite contour integral over all branches
of the analytically continued logarithm $\log(\ww)$. Thus, we
forget (the sum over) $\tmu$ in the following and understand the
contour integral as the appropriate infinite integral.

We continue with the {\it saddle point
approximation} of (2.8) (in short: the value of the integral is
approximated by the integrand at the saddle point):

\noindent
A necessary condition for a minimum in a point $\ww$ (in lowest
$\log(q)$ order, only then $\ww$ is independent of $q$, \ie a number)
is given by
$$
\eqalign{
  0 & = w_a \partial_{w_a}\ \eml(q,\ww) + {\cal O} (\log^0(q)) \cr
    & = {1\over \log(q)} \Big(
        - 2\pi i B_{aa'}^{-1}({\log(\ww)\over 2\pi i}+l\cc)_{a'}/2
        + \log(1-w_a)
     \Big) \cr }
\eqno(2.10)
$$
or
$$
  {1\over 2} B_{aa'}^{-1}({\log(\ww)\over 2\pi i}+l\cc)_{a'}
    = {\log(1-w_a)\over 2\pi i}
\eqno(2.11)
$$
or in exponentiated form (ignoring for the moment possible phases)
$$
   w_{a'} = \prod_{a=1}^n\ (1-w_a)^{2B_{aa'}},
\eqno(2.12)
$$
which appeared in the physical literature as certain limits of
TBA-type (thermodynamic Bethe Ansatz) equations. We will assume in the
following that a unique solution of (2.10,11) with $0<w_a<1$ exists and
that furthermore $l=0$ for this solution, which is indeed
true for all examples known to us.

\noindent
Substituting (2.11) into (2.8b) and introducing Rogers' dilogarithm
function
$$
     \lll(z) = \LL(z) + {1\over 2} \log(z)\log(1-z),
\eqno(2.13)
$$
we obtain the exponent function at the saddle point $\ww=\ww_s (l=0)$
(repeated indices are summed over)
$$
\eqalign{
\eml(q,\ww=\ww_s) =
&\log(\tilde q)\ \Big\{
-{1\over 24\ \lll(1)}\sum_{a=1}^n \lll (w_a) \Big\} \cr
+ & \log^0(q)\ \Big\{ \varphi_1(b_a)\
   \log (1-w_{a}) \Big\}\cr
+ & \log^1(q)\ \Big\{ \beta -
   {\varphi_2(b_a)\over 2}\ {w_a\over 1-w_a}  \Big\}
+ {\cal O} (\log^2(q)). \cr }
\eqno(2.14)
$$

The $\log(\tilde q)$ part of (2.14) yields the first and crudest
approximation of the integral. To extract {\it higher order corrections},
we change the variable of integration in (2.8) by the substitution
$w_a = e^{\sqrt{-2\pi t}\ v_a}=e^{\sqrt{\log(q)}\ v_a}$
($\partial_{v_a} =  \sqrt{\log(q)}\ w_a\partial_{w_a} $) and
expand $\eml(q,\vv)$ in a Taylor series around the saddle point
$\vv_s$ to obtain
$$
 Z(q) =  \int\
  {1\over K\sqrt{\det(2B)}}\
   \sum_{ l\in\zzz_K}\  \exp(\eml(q,\vv_s +\tD))\
     \prod_j{d\tD_j\over\sqrt{2\pi}} ,
\eqno(2.15a)
$$
where
$$
\eqalign{
\eml(q,\vv_s+\tD)=
 \eml(q,\vv) \big|_{\vv=\vv_s}
     & + \sum_a \tD_a  \partial_{\vv_a} \eml(q,\vv)
         \big|_{\vv=\vv_s}\cr
     & + {1\over 2} \sum_{a,a'}  \tD_a \tD_{a'} \ \partial_{v_a}
          \partial_{v_{a'}} \eml(q,\vv) \big|_{\vv=\vv_s}  \cr
     & + {1\over 6} \sum_a \tD_a^3 \ \partial_{v_a}^3
        \eml(q,\vv) \big|_{\vv=\vv_s}    + \ldots \quad . \cr}
\eqno(2.15b)
$$
The first derivative of $\eml(q,\vv)$ is given by
$$
\partial_{v_a} \eml(q,\vv) \big|_{\vv=\vv_s} =
  - \sum_{p=1}^{\infty}\ \log^{p-1/2}(q)\
       {\varphi_p(b_a) \over p!}\
       \sum_{r_a=1}^{\infty} w_a^{r_a} r_a^{p-1},
\eqno(2.16)
$$
the second by
$$
\eqalign{
\partial_{v_a} \partial_{v_{a'}}\
 \eml(q,\vv) \big|_{\vv=\vv_s} =
 -  & (2 B)^{-1}_{aa'} -\delta_{a,a'} {w_a\over 1-w_a} \cr
& - \delta_{a,a'}\ \sum_{p=1}^{\infty}\ \log^p(q)\
       {\varphi_p(b_a) \over p!}\
       \sum_{r_a=1}^{\infty} w_a^{r_a} r_a^{p}, \cr }
\eqno(2.17)
$$
and the higher ones for $n\geq 3$ by
$$
\partial_{v_a}^n\ \eml(q,\vv) \big|_{\vv=\vv_s} =
  - \sum_{p=0}^{\infty}\ \log^{p-1+n/2}(q)\
       {\varphi_p(b_a) \over p!}\
       \sum_{r_a=1}^{\infty} w_a^{r_a} r_a^{p+n-2}.
\eqno(2.18)
$$

The next $\log(q)$ order of our expression of $Z(q)$ can then be
found by evaluating the integral as a gaussian integral. Namely, for
any symmetric, real, positive definite matrix $A$, one has
$$
 \int\ e^{-{1\over 2} xAx^t-d\cdot x-c}\ {d^nx\over (2\pi)^{n/2}}
     = {1\over\sqrt{\det{A}}}\ e^{{1\over 2} dA^{-1}d^t-c},
\eqno(2.19)
$$
leading to
$$
\eqalign{
 Z(q) =
 & \exp \bigg( \log({\tilde q}) \Big\{
    -{1\over 24\lll(1)}\ \sum_{a=1}^n \lll (w_a) \Big\} \bigg) \cr
& \exp\Big( \sum_{a=1}^n \varphi_1(b_a)\ \log(1-w_a) \Big)
   \ \Big/ \  K\sqrt{ \det(2B\cdot A) } \cr
& \Big( 1+{\cal O}\big( \log(q) \big) \Big)   , \cr }
\eqno(2.20)
$$
where $d_a = \varphi_1(b_a) {w_a\over 1-w_a}\ \sqrt{\log(q)}$ and
$$
  A_{aa'}=(2B)_{aa'}^{-1}+\delta_{a,a'} {w_a\over 1-w_a} .
\eqno(2.21)
$$
The next order in $\log(q)$ leads to replacing $
\Big( 1+{\cal O}\big( \log(q) \big) \Big)$ by
(sums over all indices $a,a'=1,\ldots n$ implied)
$$
\eqalign{
  \exp \bigg\lbrack \log^1(q)\ \bigg(  &
       \Big\{ \beta - {\varphi_2(b_a)\over 2}\ {w_a\over 1-w_a}
        \Big\} \cr
 &  + {1\over 2}
     \varphi_1(b_a) {w_a\over 1-w_a}  A^{-1}_{aa'}
     \varphi_1(b_{a'}) {w_{a'}\over 1-w_{a'}} \cr
 &  - {1\over 2}
     \varphi_1(b_a) {w_a\over (1-w_a)^2}  A^{-1}_{aa} \cr
 &  + {1\over 2}
      {w_a\over (1-w_a)^2}  A^{-1}_{aa}  A^{-1}_{aa'}
      \varphi_1(b_{a'}) {w_{a'}\over 1-w_{a'}} \cr
 &  - {1\over 8}
     {w_a(1+w_a) \over (1-w_a)^3}  (A^{-1}_{aa})^2 \cr
 &  + {1\over 12}
      {w_a\over (1-w_a)^2} (A^{-1}_{aa'})^3
     {w_{a'}\over (1-w_{a'})^2} \cr
 &  + {1\over 8}
      {w_a\over (1-w_a)^2} A^{-1}_{aa} A^{-1}_{aa'} A^{-1}_{a'a'}
     {w_{a'}\over (1-w_{a'})^2}
       \bigg) \bigg\rbrack \cr
 & \phantom{mmmmmmmmmmm}
   \Big( 1 + {\cal O}\big(\log^2(q)\big) \Big) \cr }
\eqno(2.22)
$$
in (2.20). It can be found by taking derivatives of (2.19) with
respect to $d$ and then collecting all terms of order $\log(q)$.
Once written down, one also recognizes it as the exponential
of the generating function of the connected Feynman diagrams of
order $\log(q)$ (with corresponding factors of internal symmetry)
allowed by the Feynman rules encoded in (2.15).

\noindent
This insight allows one to write down the next order by hand
(with $35$ 'new' diagrams, meaning diagrams not constructed out of
'dressings' ($p\mapsto p+1$ in (2.16-18)) of (2.22)).
The next to next order already is a bit unwieldly with
367 diagrams before choosing possible dressings.
Higher orders can in principle be written down. However, the
proliferation of diagrams is enormous and we have looked
beyond $\log(q)$ only in the comparably simple case where
${\rm rank}\ B=1$.

\medskip 
\unter 3 Modular covariance

In this section we compare the asymptotics found above with the
prediction of modular covariance.

In the following, we suppose that the partition $Z(q)$ given in
$(1.1)$ equals a character of some CFT with effective central charge
$\cef = c-24\ \hm$ (the index $\min$ refers to the primary field
of minimal conformal dimension) -- more accurately: $Z(q)$ equals
the character corresponding to the primary field $\phi_{\lambda}$
of conformal dimension $h_{\lambda}$, thus
$$
   Z^{(\lambda)}(q)=
   q^{h_{\lambda}-c/24}\sum_{m\in\zzz_{\geq 0}}d^{(\lambda)}_m q^m .
\eqno(3.1)
$$
Then by modular covariance we have
$$
  Z^{(\lambda)}(q)
  =\sum_{{\lambda}'} S_{{\lambda},{\lambda}'}\
       Z^{(\lambda')}(\tilde q)
  =\sum_{{\lambda}'} S_{{\lambda},{\lambda}'}\
          {\tilde q}\, {}^{h_{{\lambda}'}-c/24}
         \sum_{m\in\zzz_{\geq 0}}d^{(\lambda')}_m\ {\tilde q}\, {}^m,
\eqno(3.2)
$$
where the $d_m^{(\lambda)}$ are the multiplicities of states at grade $m$.

\noindent
Comparing (1.1) with (3.1), we see that
$$
h_{\lambda}-c/24=
\min_{\bf m\geq 0}({\bf m}B{\bf m}^t + {\bf b\cdot m} + \beta).
\eqno(3.3)
$$

\noindent
Comparing the first term (in powers of ${\tilde q}$) of the right hand
side of (3.2) with (2.20,22), we obtain
$$
\eqalignno{
& \cef =  \sum_{a=1}^n \ {\lll (w_a)\over \lll (1)} &(A)\cr
&  S_{\lambda ,\min} = {\exp\Big(
   \sum_{a=1}^n \varphi_1(b^{(\lambda)}_a)\
   \log(1-w_a) \Big) \over
   K \sqrt{ \det(2B\cdot A)} } &(B)\cr
& 0 = {\rm exponent\ of\ (2.22)}
  \big|_{\beta=\beta^{(\lambda)},\bb=\bb^{(\lambda)}}
  &(C)\cr
& 0 = {\rm higher\ orders\ in\ \log(q)\ of\ the\
        expansion\ of\ } Z^{(\lambda)}(q), &(D)
}
$$
where as before
$$
  A_{aa'}=(2B)_{aa'}^{-1}+\delta_{a,a'}
                    {w_a\over 1-w_a} .
$$
In words:

\noindent
$(A)$ gives the effective central charge of the theory in
terms of a dilogarithmic expression with arguments $w_a$ given
by a certain solution of the TBA-type equations (2.12) which only
depend on $B$.

\noindent
$(B)$ gives an expression for certain elements of the modular
S-matrix in terms of the $w_a$.
{}From $(B)$ we obtain the following relation between
quantum dimensions and solutions of the TBA-type equations
$$
  { S_{\lambda ,\min}\over S_{0 ,\min}}
  =  \ \prod_{a=1}^n\
  (1-w_a)^{\displaystyle
  b_a^{(\lambda)}-b_a^{(0)}} ,
\eqno(3.4)
$$
where $0$ denotes the vacuum of the theory.

\noindent
$(C)$ fixes $\beta$ in terms of $w_a$ (\ie $B$) and $\bb$, thus
with (3.3) gives an expression for the conformal dimensions
of the theory.

\noindent
Finally, $(D)$ is an innocent looking short-hand for an infinite set
of complicated consistency conditions for modular covariance
putting constraints on $B$  and
$\bb$. However, trying to solve them is highly non trivial and of the
type to solve $(C)$ for given $\beta$ in terms of $B$ and $\bb$. For
the case ${\rm rank}\ B=1$ we refer the reader to the outlook.

\medskip 
\unter 4 Outlook

Assuming the existence of a character with $b=0$, we have classified
the partitions of type (1.1) with rank $B=n=1$:

\noindent
We found numerically the zeroes of the first non-trivial consistency
condition (order $\log^2(q)$) in $w\in\rbrack 0,1\lbrack$, namely
$\tau,\ 1/2,\ \tau^2$ ($\tau^2+\tau-1=0$), which correspond to the
$(3,5)-$, $(3,4)-$ and $(2,5)-$model in the series of Virasoro
minimal models, with $B=1/4,1/2,1$ respectively.
Given $w$ and thus $B$, we could solve the first consistency
conditions for other possible $b\neq 0$ and found the known results.

Certainly one should try to attack the higher rank case by the same
method and to extend the argument to the more general partitions
mentioned in the introduction.

\medskip
Also, we would like to understand, in a way similar to the discussion
given, other dilogarithm identities leading to the other effective
conformal dimensions or possibly (\cf Kuniba, Nakanishi, Suzuki (1992))
to the whole spectrum of CFTs, \ie we would like to be able to
explicitly construct the whole modular transform of partitions in
the Rogers-Ramanujan form and not only the first term. A 'naive'
analytic continuation of the (di)logarithm in (2.20) leads to a form
surprisingly similar to (1.1) again. However, a closer lock reveals
certain inconsistencies and substantially reduces the predictive power
(only $(3.C)$ seems to hold without any 'tuning').
This problem is solved by W. Nahm (1993).

\medskip 
\unter A \kern-6pt cknowledgements

I thank W. Nahm for encouragement and advice.

I am indebted to W. Eholzer, R. Kellerhals, Andi Recknagel, M. R\"osgen,
R. Varnhagen and D. Zagier for essential discussions on related
subjects.

During the time of writing I was supported by my parents and
the Studien\-stiftung des deutschen Volkes.

\medskip 
\unter R \kern-6pt eferences

\baselineskip=12pt

\frenchspacing

\bigskip
\parindent=-1truecm
\parskip=0.2cm
\leftskip=1truecm

G.E. Andrews (1976), {\sl The theory of partitions}, Encyclopedia
   of Mathematics and Its Applications, Vol. 2, Addison Wesley.

V. V. Bazahnov, N. Yu. Reshetikhin (1989), {\sl Int. J. Mod. Phys. A}
   {\bf 4}, 115.

H. M. Babujan (1983), {\sl Nucl. Phys.} {\bf B215}, 317-36.

S. Dasmahapatra, R. Kedem, T. R. Klassen, B. M. McCoy, E. Melzer (1993),
   {\sl Quasi-Particles, Conformal Field Theory, and q-Series},
   ITP-SB-93-12, RU-93-07, hep-th/9303013, to appear in the Proc. of
   Yang-Baxter Equations in Paris Conf., Paris, France, July 24-30,
   (1992?).

B. Davies (1978), {\sl Integral transforms and their applications},
    Applied mathematical sciences, Vol. 25, Springer-Verlag.

R. C. Gunning (1962), {Lectures on Modular Forms}, Princeton
    University Press.

T. R. Klassen, E. Melzer (1990), {\sl Nucl. Phys. }{\bf B338}, 485-528;
  (1992), {\sl Nucl. Phys. }{\bf B370}, 511.

A. Kl\"umper, P. A. Pearce (1991), {\sl J. Stat. Phys.} {\bf 64}, 13.

A. Kuniba, T. Nakanishi (1992), {\sl Mod. Phys. Lett.}
        {\bf A7} (1992), 3487-94.

A. Kuniba, T. Nakanishi, J. Suzuki (1992),
    {Characters in CFTs from Thermodynamic Bethe Ansatz},
     preprint hep-th/9301018, HUTP-92/A069.

G. Meinardus (1954), {\sl Math. Zeitschr. }, Bd. {\bf 1}, 289-302.

W. Nahm, A. Recknagel, M. Terhoeven (1992), hep-th/9211034,
   preprint BONN-HE-92-35, to appear in {\sl Mod. Phys. Lett.} {\bf A}.

W. Nahm (1993), in preparation.

B. Richmond, G. Szekeres (1981),
   {\sl J. Austral. Math. Soc.} {\sl (Series A) 31}, 362-373.

M. Terhoeven (1992), {\sl Lift of dilogarithm to partition identities},
preprint BONN-HE-92-36, hep-th/9211120.

Al. B. Zamolodchikov (1990), {\sl Nucl. Phys. B} {\bf 342}, 695.

\eject
\bye